\documentclass[twocolumn, notitlepage, superscriptaddress, aps, longbibliography]{revtex4-1}
\usepackage{color}
\usepackage{amsmath}
\usepackage{amsthm}
\usepackage{graphicx}
\usepackage{float}
\usepackage{hyperref}
\usepackage{ulem}
\usepackage{caption}
\usepackage[position=top]{subfig}
\usepackage{framed}
\newcommand{\supp}{\color{blue} Supplement 1 \color{black} }
\graphicspath{{Figures/}}
\newcommand{\pd}[2]{\frac{\partial #1}{\partial #2}}

\newcommand{\ket}[1]{\left| #1 \right\rangle}
\newcommand{\bra}[1]{\left\langle #1 \right|}

\newcommand{\dg}{^\dagger}
\newcommand{\omg}{\omega}

\begin{document}

\title{Enhancing LIDAR performance metrics using continuous-wave photon-pair sources}

\author{Han Liu}
\author{Daniel Giovannini}
\affiliation{The Edward S.~Rogers Department of Electrical and Computer Engineering, University of Toronto, 10 King's College Road, Toronto, Ontario M5S 3G4, Canada}
\author{Haoyu He}
\affiliation{The Edward S.~Rogers Department of Electrical and Computer Engineering, University of Toronto, 10 King's College Road, Toronto, Ontario M5S 3G4, Canada}
\author{Duncan England}
\affiliation{National Research Council of Canada, 100 Sussex Drive, Ottawa, Ontario, K1A 0R6, Canada}
\author{Benjamin J. Sussman}
\affiliation{National Research Council of Canada, 100 Sussex Drive, Ottawa, Ontario, K1A 0R6, Canada}
\affiliation{Department of Physics, University of Ottawa, 598 King Edward, Ottawa, Ontario K1N 6N5, Canada}
\author{Bhashyam Balaji}
\affiliation{Radar Sensing and Exploitation Section, Defense R\&D Canada, Ottawa Research Centre, 3701 Carling Avenue, Ottawa, Ontario, K1A 0Z4, Canada}
\author{Amr S.~Helmy}
\email{a.helmy@utoronto.ca}
\affiliation{The Edward S.~Rogers Department of Electrical and Computer Engineering, University of Toronto, 10 King's College Road, Toronto, Ontario M5S 3G4, Canada}
\begin{abstract}
In order to enhance LIDAR performance metrics such as target detection sensitivity, noise resilience and ranging accuracy, we exploit the strong temporal correlation within the photon pairs generated in continuous-wave pumped semiconductor waveguides.
The enhancement attained through the use of such non-classical sources is measured and compared to a corresponding target detection scheme based on simple photon-counting detection.
The performances of both schemes are quantified by the estimation uncertainty and Fisher information of the probe photon transmission, which is a widely adopted sensing figure of merit.
The target detection experiments are conducted with high probe channel loss (\(\simeq 1-5\times10^{-5}\)) and formidable environment noise up to 36 dB stronger than the detected probe power of \(1.64\times 10^{-5}\) pW.
The experimental result shows significant advantages offered by the enhanced scheme with up to 26.3 dB higher performance in terms of estimation uncertainty, which is equivalent to a reduction of target detection time by a factor of 430 or 146 (21.6 dB) times more resilience to noise.
We also experimentally demonstrated ranging with these non-classical photon pairs generated with continuous-wave pump in the presence of strong noise and loss, achieving \(\approx\)5 cm distance resolution that is limited by the temporal resolution of the detectors.
\end{abstract}

\maketitle
\setcounter{tocdepth}{1}

\section{Introduction}
Non-classical attributes of light have been utilized to extend the sensitivity of several metrology applications beyond the classical limit recently\cite{aasi2013enhanced}\cite{brida2010experimental}\cite{losero2018unbiased}. 
One important sensing technique; namely optical target detection, can benefit greatly from non-classical correlations of photons.
Optical target detection has been receiving increasing attention due to its pivotal role in emerging applications such as light based detection and ranging (LIDAR)\cite{takai2016single} and non-invasive bio-imaging, among others.
Quantum protocols based on quantum illumination have recently demonstrated that entanglement can be utilized to beat the best classical target detection schemes\cite{lloyd2008enhanced}.
The enhancement obtained from such schemes requires phase-sensitive quantum detection, which entails that the optical path length of the probe light has to be stabilized precisely, on a wavelength scale, while the reference photons have to be stored in a quantum memory\cite{Zhang:2015}.
Such stringent requirements preclude, at this stage, quantum illumination-based schemes from addressing numerous practical sensing and ranging application.
In contrast, a phase-insensitive target detection scheme which utilizes non-classical primitives, if proven to provide enhancement over its classical counterparts, would be a valuable advance in the quest to improve upon classical limits of detection.
When developed, such a scheme can be beneficial for many practical sensing and ranging applications.

The essential idea of nonclassical light enhanced phase-insensitive target detection scheme is to utilize the correlation within nonclassical photon pairs to enhance the target detection performance. Previous implementations of such schemes \cite{Lopaeva:2013}\cite{england2019quantum} are based on the intensity correlation within non-classical photon pairs, that is, the correlation between the number of probe and reference photons. However, the quantum enhancement of intensity correlation-based schemes diminishes as the output power of the probe light increases\cite{england2019quantum}, which limits its application in practical target detection scenarios. Another type of correlation that also exists within non-classical photon pairs is the temporal correlation, that is, the correlation between the detection time of the probe and reference photons. Compared to the intensity correlation, the temporal correlation is not limited by the source power and is therefore much more scalable. As such, it can be interesting to see whether the temporal correlation could be utilized to benefit practical phase-insensitive target detection schemes. 

One of the central aspects of practical target detection systems is the versatility of the non-classical source used. To date, non-classical photon sources for most quantum-enhanced metrology experiments have been based on bulk crystal such as BBO and PPLN with a large footprint, which inevitably introduces mechanical instability and inefficient interaction with the metrology system.
For practical target detection schemes utilizing non-classical photon pairs, the sources need to be in spatial single-mode and offer a compact footprint, to cater to the widest range of applications.
Integrated photon-pair sources are therefore ideal candidates for these applications.
In the last decade, there has been astounding progress in the prowess of non-classical sources\cite{Kang:2016}\cite{Kang:2015}.
In particular, it has been shown that integrated monolithic semiconductor devices based on an active gallium arsenide (GaAs) platform can be used to generate high-quality quantum states of light \cite{Horn:2012}\cite{Valles:2013}.
In addition to its compact footprint as well as efficient coupling into integrated metrology platforms, monolithic semiconductor photon-pair sources could also offer great tuning of their spectro-temporal properties electro-optically with no moving parts, which is not available in bulk photon-pair sources.
For example, the spectral bandwidth of the non-classical photon pairs could be tailored from 1 nm to over 450nm through varying the waveguide structure with various mask designs\cite{Abolghasem:2009}.\\

Another aspect of paramount importance is to choose a suitable criterion, based on which a meaningful and practical figure of merit (FOM) can be defined to facilitate comparisons between classical and non-classical schemes.
In previous experiments\cite{Lopaeva:2013}\cite{england2019quantum}, the target detection performance is quantified by using the optical signal to noise ratio (OSNR).
The calculation of OSNR necessitates separate target detection experiments with the target being present and absent, which may not be practical in applications where the removal of the target is not feasible.
Moreover, the definition of OSNR may vary depending on the implementation, rendering performance comparisons challenging to standardize.
A robust approach to address these issues is to adopt the estimation theory, which has been widely used and proven as a reliable figure of merit for various target detection and sensing applications\cite{sanz2017quantum}\cite{whittaker2017absorption}.
If the estimate of the probe light transmission, which is defined as the percentage of probe photons that are back-reflected from the target and get detected, is greater than the uncertainty of estimation, then the probability of the presence of the target object could be maximized.
The uncertainty of the transmission estimation, which could be obtained through a single target detection experiment, can serve as an effective, and generic, target detection FOM. This is because it characterizes the ability to distinguish between the presence (nonzero probe light transmission) and absence (zero probe light transmission) of the target object.
The uncertainty of the transmission estimation is lower bounded by the Fisher information of the transmission estimation according to the Cram\`er Rao's Bound\cite{cramer1999mathematical}.\\


In this work, we demonstrate a practical, phase-insensitive target detection scheme which utilizes photon pairs obtained from a continuous-wave (CW) pumped spontaneous parametric down-conversion (SPDC) source realized using a monolithic semiconductor waveguide.
We show that the strong temporal correlation of the non-classical photon pairs could provide substantial and scalable performance enhancement to phase-insensitive target detection schemes.
The classical target detection system, which we use as a comparison to the non-classical sources enhanced scheme, utilizes simple intensity detection with the same photon-counting detector that is used in the enhanced scheme.
The performances of both schemes are quantified by using the transmission estimation uncertainty and Fisher information criterion.
Furthermore, we demonstrate that temporal correlation enhanced systems are also resilient to active jamming attacks and are capable of effective ranging in a noisy and lossy environment, despite the utilization of CW light and not pulsed.
 The experimental results match well with the theoretical predictions obtained from the Fisher information and the Cram\'er-Rao's Bound.  \\
\onecolumngrid
\begin{center}
\begin{figure}[h]
    \centering
    \includegraphics[width=0.8\columnwidth]{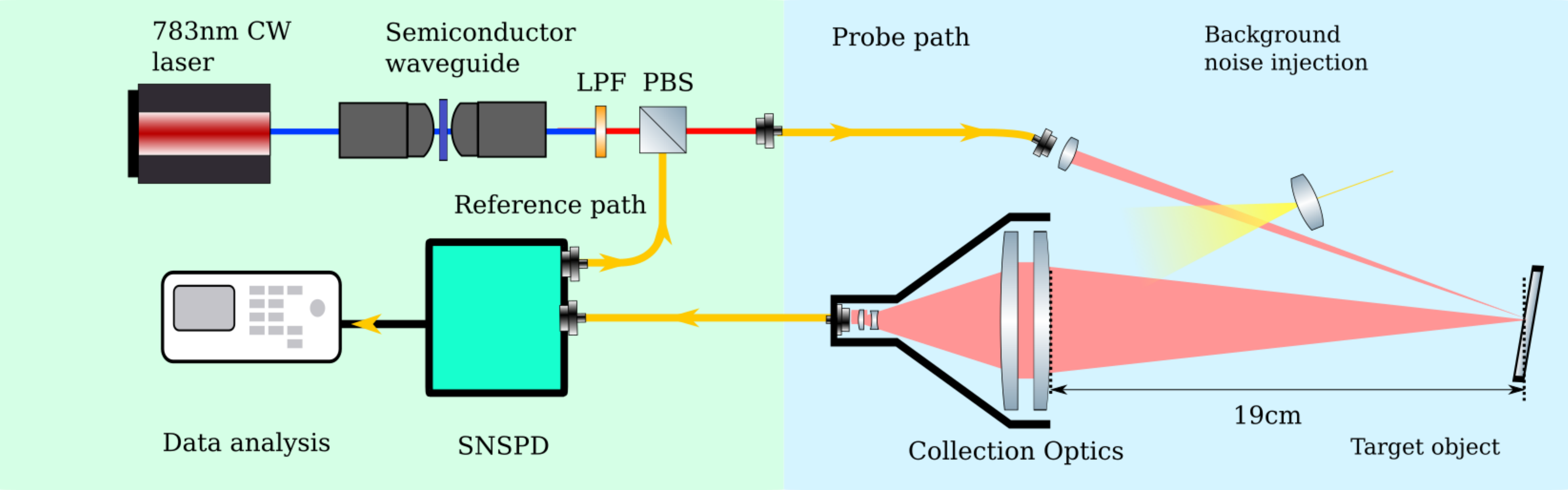}
    \caption{The schematic of the experimental setup which is divided into two parts. The left part (green background) includes the probe photon source and detectors. Pump laser: Ti-Sapphire CW laser at 783nm. PBS: polarization beam-splitter. LPF: long-pass (\(>\)1200 nm) filter to separate the SPDC photon pairs from the pump laser. SNSPD: dual-channel superconducting single-photon detector (top channel: the reference detector, bottom channel: the probe detector). Right part (blue background): probing of the target object and collection of the scattered photons. Target object: a piece of white paper with diffused reflection. The left and right part are built on separate tables and are connected by single-mode fibers (yellow line with arrows).  }\label{setup}
\end{figure}
\end{center}
\twocolumngrid

\section{Phase-insensitive target detection using non-classical light}
In this section, we theoretically analyze the phase-insensitive target detection scheme using non-classical photon pairs.
We compare its performance to a classical target detection scheme with intensity (photon counting) detection.
We quantify the performance of both the classical and non-classical target detection schemes using transmission estimation uncertainty and Fisher information as the FOM.
In particular, we show that the enhancement obtained using non-classical light is directly related to the strong temporal correlation of the source used.\\

We consider a phase-insensitive target detection scheme enhanced by non-classical photon pairs (probe and reference photons) that are generated in the semiconductor waveguide. The scheme will be called classical target detection using non-classical photons, or CDNC in this article.
The SPDC process is pumped by a CW laser to obtain strong temporal correlation between the probe and reference photons.
For each generated photon pair, the reference photon is immediately detected on a single-photon detector (the reference detector) and the probe photon is directed towards the unknown object.
If the object is present, a small fraction of probe photons are reflected from the object impinging and detected with a finite efficiency at the second single-photon detector (the probe detector).
A controllable level of background noise is always coupled onto the probe detector regardless of the presence of the object.
We assume that the probe photons are indistinguishable from the background noise photons in terms of their `local properties', e.g., temporal pulse shape and spectral distribution. Therefore it is not possible to reduce the noise power through any classical filtering technique (i.e. temporal gating, spectral filtering or mode selection).
The photon detection statistics from both detectors are recorded for estimating the probe channel transmission, which is defined as the percentage of probe photons (not including noise photons) that get back-reflected and detected on the probe detector.
If the estimated channel transmission is greater than the estimation uncertainty, the likelihood of the presence of the unknown object increases proportionally.
Therefore the uncertainty of the transmission estimation characterizes the sensitivity of this target detection system.
For comparison, we consider a classical phase-insensitive target detection scheme using classical photons, which will be called, CDC in this text. CDC utilizes simple photon counting, with no coincidences, as a means of detection.
The classical probe photons are sent towards the object. The back-reflected photons from the object (none if the object is not present), together with the background noise photons are detected and recorded on a single-photon detector.
Same as the CDNC scheme, the likelihood of the presence of the unknown object is increased proportionally to the estimated channel transmission increase beyond the estimation uncertainty.
Note that the CDC scheme can be considered as a special case of the CDNC scheme, where all the reference photons are blocked.
Therefore the same analysis of the CDNC scheme applies.
This approach represents a more general and straight-forward classical phase-insensitive target detection scheme compared to the previous schemes based on classical intensity correlation of thermal light\cite{Lopaeva:2013}. The SPDC photon pair state generated in the waveguide could be expressed as:
\begin{gather}
\ket{\Psi} = \ket{vac}+  \int_{-\infty}^{\infty}\int_{-\infty}^{\infty}\phi(t_p-t_r)\notag\\
\exp(-i\omg_{p,0} t_p-i\omg_{r,0} t_r )a_p\dg(t_p)a_r\dg(t_r)dt_pdt_r\ket{vac}
\end{gather}
where \(\phi(t_p-t_r)\exp(-i\omg_{p,0} t_p-i\omg_{r,0} t_r )\) is the joint temporal amplitude and \(\omg_{p,0},\omg_{r,0}\) are the central frequencies of the probe and reference photons.   
The creation operator of the probe (reference) photon at time \(t\) is denoted by \(a_p\dg(t)\) (\(a\dg_r(t)\)).
The intrinsic temporal correlation time \(\Delta t_0\) of the SPDC photon pairs are defined as the standard deviation of the detection time difference between the probe and reference photons (assuming infinite detection temporal resolution): 
\begin{gather}
\Delta t_0 = (\int_{-\infty}^{+\infty} dt_p \frac{|\phi(t_p-t_r)|^2}{\int_{-\infty}^{+\infty}|\phi(t_p'-t_r)|^2 dt_p'}(t_p-t_r)^2)^\frac{1}{2}
\end{gather}
It could be shown that \(\Delta t_0\) is inversely proportional to the SPDC photon bandwidth\cite{franson1992nonlocal}. In our implementation the bandwidth is over 100 nm, which corresponds to \(\Delta t_0 = 35\)fs(see \supp for the characterization details). 
The output flux \(\nu\) of the probe photons from the source is given by\cite{huttner1990quantum}:
\begin{gather}
\nu = tr\{a_p\dg(t_p)a_p(t_p)\ket{\Psi}\bra{\Psi}\}=\int_{-\infty}^{+\infty}|\phi(t_p-t_r)|^2dt_r
\end{gather}

where \(tr\) stands for the trace over the joint Hilbert space of the reference and probe photons.
Note that \(\nu\) is independent of time \(t_p\) because the SPDC process is pumped using a CW laser.
The output flux of the reference photons is also given by \(\nu\) since the probe and reference photons are always generated in pairs.

The total transmission efficiency of the reference photons \(\eta_r\) is affected by the optical coupling loss and the inefficiency of the reference detector.
The total transmission efficiency of the probe photons \(\eta_p\) is also affected by the reflectivity of the target object and the collection efficiency of the back-reflected probe photons.
The flux of the noise photons that get detected on the probe detector is denoted by \(\nu_b\).
Assuming the target detection system to be time-invariant, i.e. stationary target, the absolute time of photon detection events carries no information about the presence of the object.
The only photon detection statistics of interest is the photon detection rate upon each detector as well as the coincidence detection rate across both detectors. A coincidence detection event is defined as a detection event on the reference detector at time \(t_r\) being followed by another detection event on the probe detector at time \(t_p\), such that \(t_p-t_r\) lies within a small coincidence window:
\begin{gather}
t_p-t_r \in [(l_p-l_r)/c-\frac{T_c}{2},(l_p-l_r)/c+\frac{T_c}{2}]\label{DEF_NC}
\end{gather}
where \(T_c\) is the temporal width of the coincidence window and \(l_p, l_r\) are the optical path length of the probe and reference photons.
To ensure that every probe-reference photon pair (not including noise-reference photon pairs) that is detected on the detectors could contribute to a coincidence detection event, the coincidence windows \(T_c\) is taken to be larger than the effective temporal correlation time \(\Delta t_{eff}\): 
\begin{gather}
T_c \ge \Delta t_{eff} = 2\Delta t+3\Delta t_0
\end{gather}
where \(\Delta t\) is the temporal resolution of both the probe and reference detectors. Then it could be shown that the coincidence detection rate is given by (see the \supp for the detailed derivation):  \\  
\begin{gather}
P_c =  \eta_r\eta_p \nu+\eta_r\nu_b\nu T_c\label{eq2}
\end{gather}
Define the rate of photon detection events on the probe(reference) detector that do not contribute to coincidence detection as \(P_p(P_r)\), then (see the \supp for the detailed derivation):
\begin{gather}
P_r =  \eta_r \nu - P_c\hspace{20pt}
P_p = \nu_b+ \eta_p\nu - P_c \label{eq1}
\end{gather}

The uncertainty of transmission estimation in the CDNC scheme and the CDC scheme could be quantified using the classical estimation theory.
We first need to model the probabilistic distribution of different photon detection outcomes for an experiment that lasts for a time duration of \(\tau\).
Let \(N_p(N_r)\) denote the number of photon detection events on the probe (reference) detector that does not contribute to coincidence detection and let \(N_c\) denote the number of coincidence detection events.
Because photon detection events at different time are independent, we assume \(N_p\),\(N_r\) and \(N_c\) follow Poisson distributions.
Then the joint probability distribution \(p(N_p,N_r,N_c;\tau)\) could be calculated from \eqref{eq2}\eqref{eq1}:
\begin{gather}
p(N_p,N_r,N_c;\tau) = f(N_p,P_p\tau)f(N_r,P_r\tau)f(N_c,P_c\tau)
\end{gather}
where \(f(N,\lambda) =\exp(-\lambda)\frac{\lambda^N}{N!} \) is the Poisson distribution function.
The overall transmission of the probe photons \(\eta_p\), which parametrize the joint probability distribution \(p(N_p,N_r,N_c;\tau)\), could be estimated from the actual photon detection statistics \(N_p,N_r,N_c\) through maximizing the probability \(p(N_p,N_r,N_c;\tau)\) by varying value of \(\eta_p\) (the maximal likelihood estimation\cite{cramer1999mathematical}).
The uncertainty of the transmission estimation is related to the Fisher information \(I\) of \(\eta_p\) contained in the probability distribution \(p(N_p,N_r,N_c;\tau)\), which is defined as:
\begin{gather}
    I = \sum\limits_{N_p,N_r,N_c=0}^{+\infty}p(N_p,N_r,N_c;\tau)(\pd{}{\eta_p}\text{log}p(N_p,N_r,N_c;\tau))^2\label{FI}\\
      = (\frac{\eta_r^2\nu^2}{P_c}+\frac{(1-\eta_r)^2\nu^2}{P_p}+\frac{\eta_r^2\nu^2}{P_r})\tau\label{FI_EXP}
\end{gather}
Note that \(I\) has a removable singularity at \(\eta_r=0\), which corresponds to the CDC scheme. 
Denote the maximal likelihood estimation of probe transmission as \(\hat{\eta}_p\).
We choose to quantify the transmission estimation uncertainty as the variance of transmission estimation. Then the total Fisher information \(I\) is related to the estimation variance \(\Delta^2\hat{\eta}_p\) according to the Cram\'er-Rao's Bound\cite{cramer1999mathematical}:
\begin{gather}
\Delta^2\hat{\eta}_p \ge  \frac{1}{I}
\end{gather}
This minimal uncertainty is asymptotically achievable for maximal likelihood estimation of \(\eta_p\) in the limit of a large number of repeated experiments\cite{ly2017tutorial}. Since the Fisher information \(I\) scales linearly with respect to experiment duration \(\tau\), the minimal estimation variance is inversely proportional to the detection time \(\tau\).
The transmission estimation criterion is a general figure of merit for phase-insensitive target detection since it could be applied to any target detection scheme and requires no prior information about the target object.\\

The performance enhancement of the CDNC scheme over the CDC scheme originates from the strong temporal correlation of the non-classical photon pairs. 
This could be seen in the explicit expression of Fisher information \eqref{FI_EXP}.
The first term in the expression represents the contribution of coincidence detection \(P_c\) to the total Fisher information \(I\) and the last two terms represent the contribution of \(P_p\) and \(P_r\), as indicated by their respective denominators.
In the presence of strong background noise and a strong loss of the probe photons, the contribution of coincidence detection dominates in the total Fisher information.
This is because noise photons will significantly increase \(P_p\) but its influence on the coincidence detection rate \(P_c\) is limited by the small coincidence window \(T_c\).
In another word, the noise resilience of the CDNC scheme (which is defined as the maximal noise power \(\nu_b\) that can be tolerated to achieve the same target detection performance) scales inversely with respect to \(T_c\), as could be seen in \eqref{eq1} and \eqref{FI_EXP}. Note that the minimal coincidence window \(T_c\), or the effective temporal correlation time \(\Delta t_{eff}\), is dictated by the detector temporal resolution \(\Delta t\) and the intrinsic temporal correlation time \(\Delta t_0\).\\

The ability of ranging is another advantage offered by the CDNC scheme, which utilizes the strong temporal correlation of the non-classical photon pairs.
This could be seen from the fact that the coincidence detection histogram(\(N_c\) versus \(t_p-t_r\)) that is recorded during the experiment will peak at the position that corresponds to the time of flight difference between the probe photon and the reference photon.
To be more specific, in the CDNC scheme, the number of coincidence detection events \(N_c\) is a function of \(l_p\) (the hypothetical target distance) as could be seen in \eqref{DEF_NC}.
Since \(N_c(l_p)\) will peak around the actual value of probe path length, so will the estimated transmission \(\hat{\eta}_p\) as a function of \(l_p\).
As a consequence, the actual target distance could be estimated from the peak position of \(\hat{\eta}_p(l_p)\) during the post data processing.
In comparison, ranging is not possible using CW sources in CDC schemes, since the detected signal \(N_p\) contains no information about the target distance.\\

\section{Experiment and result}
The experimental setup of the CDNC scheme is shown in Fig. (\ref{setup}).
The GaAs waveguide source is based on a Bragg-AlGaAs structure \cite{Horn:2012} which has a ridge width of 5 \(\mu\)m and length of 1 mm.
The Bragg mode of the waveguide is pumped using a CW source with a wavelength of 783 nm.
It generates photon pairs with 1566nm central wavelength through the type II SPDC process. 
The SPDC conversion efficiency is estimated to be 2.1\(\times 10^{-8}\) (photon pairs/pump photon) \cite{Horn:2012}.
\begin{figure*}[htbp]
\centering
    \subfloat[]{\includegraphics[width=0.85\columnwidth]{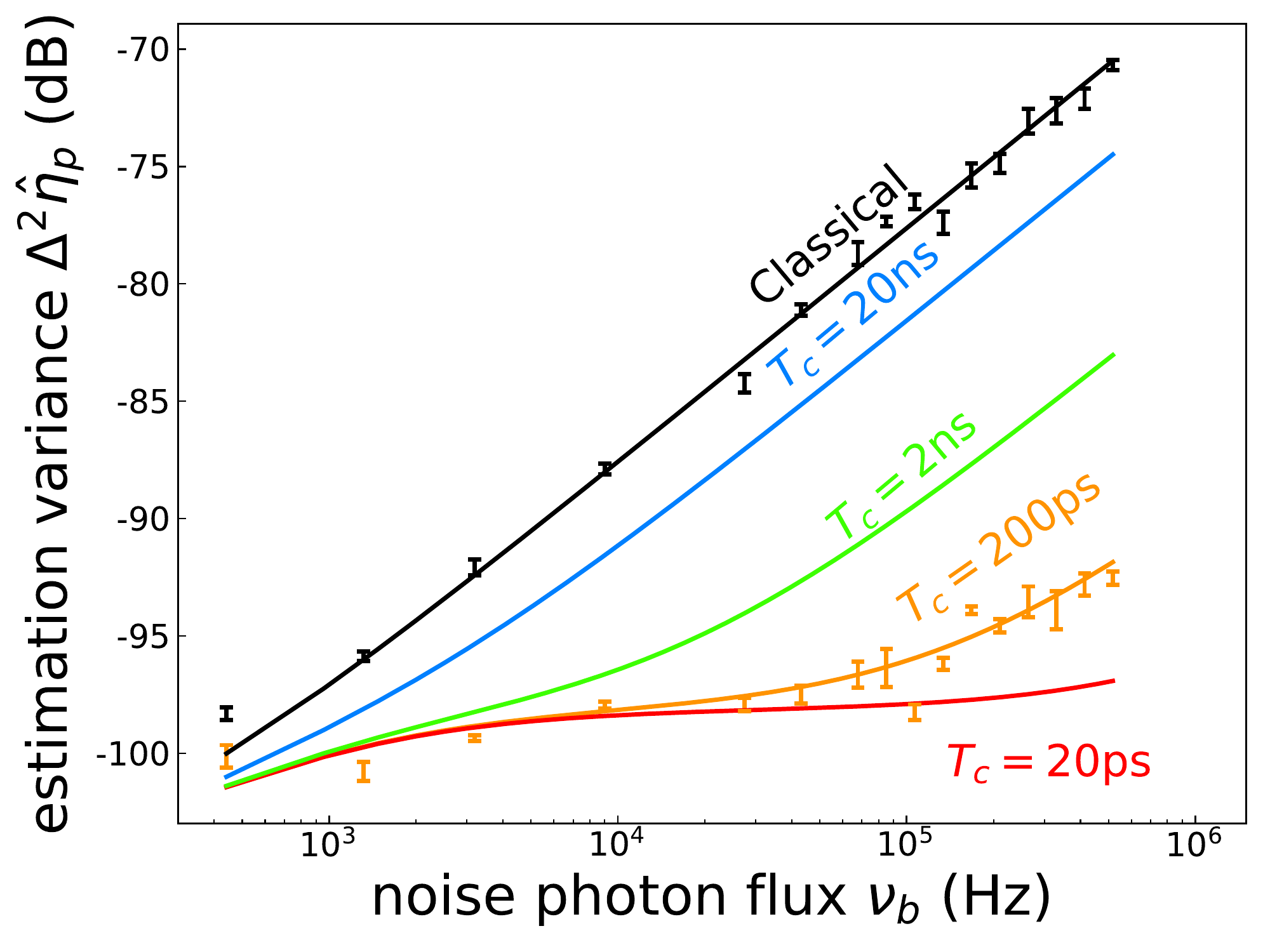}\label{n}}  
    \subfloat[]{\includegraphics[width=0.85\columnwidth]{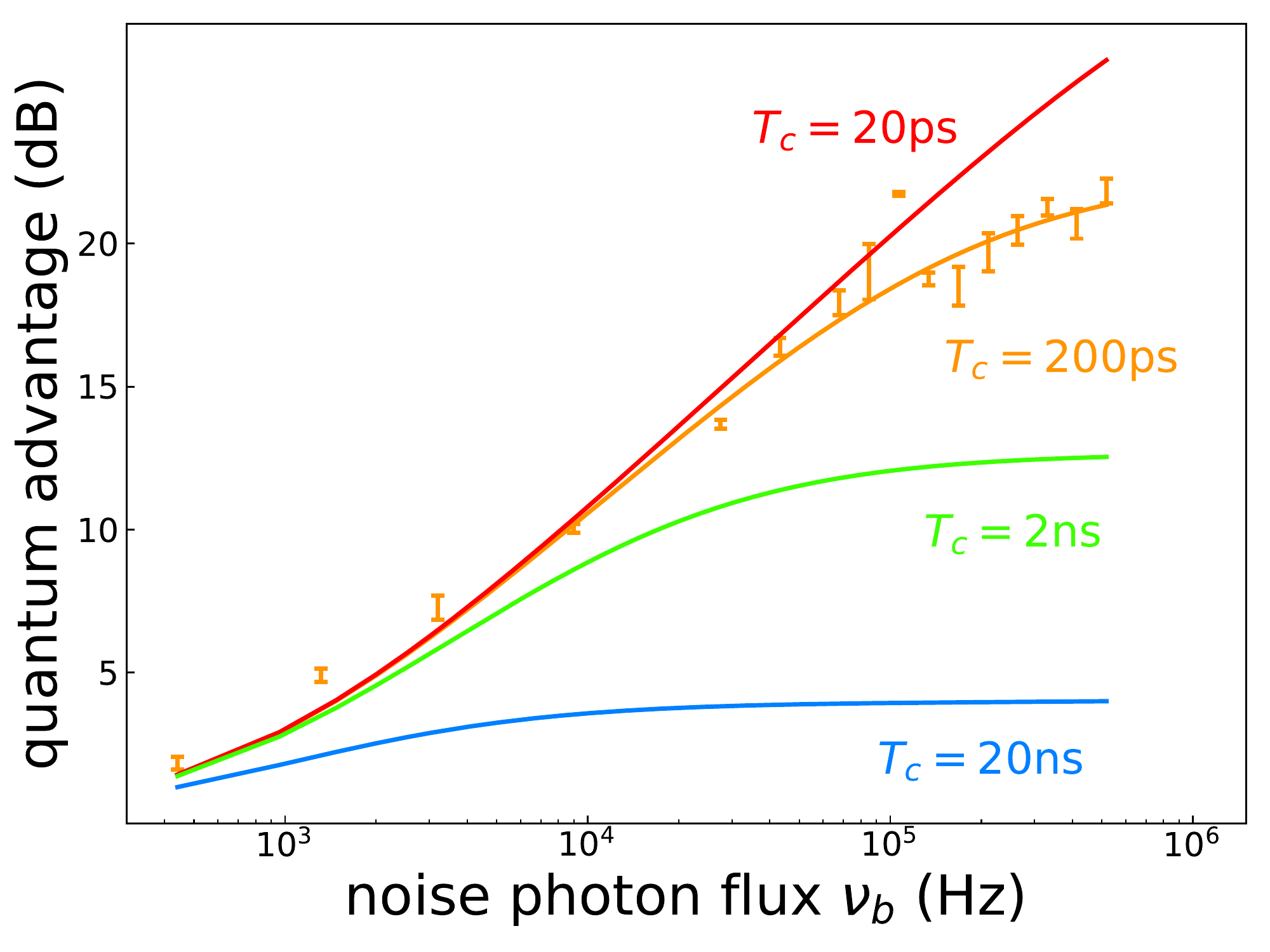}\label{nr}}\\
    \subfloat[]{\includegraphics[width=0.85\columnwidth]{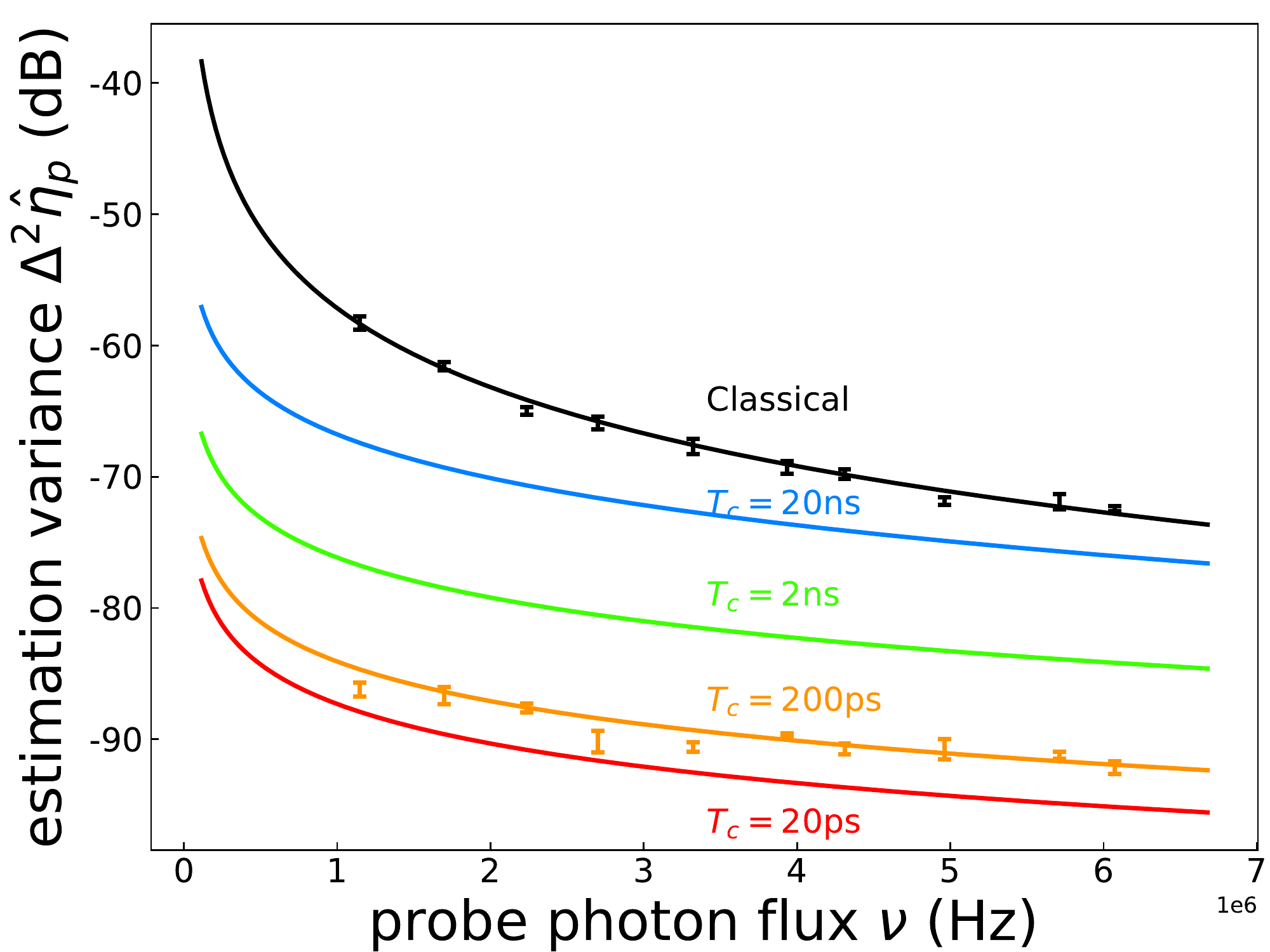}\label{p}}   
    \subfloat[]{\includegraphics[width=0.85\columnwidth]{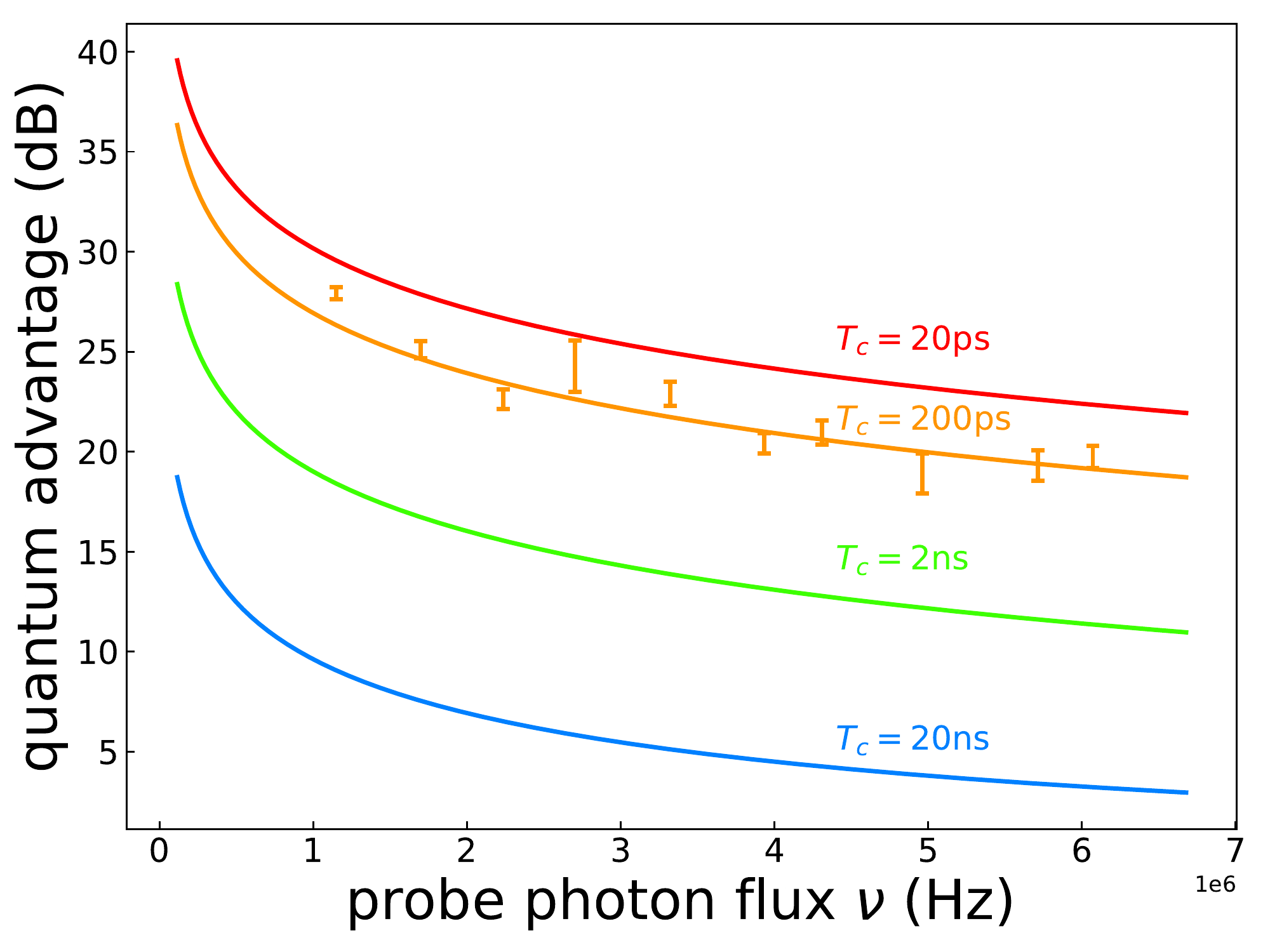}\label{pr}}   
    \caption{(a) Experimentally measured (error bar) and theoretically predicted (solid line) estimation variance \(\Delta^2 \hat{\eta}_p\) of the CDNC (orange) and CDC (black) scheme as a function of the noise flux \(\nu_b\) with \(\eta_p = 3.3\times10^{-5}\) and \(\eta_r =17.8\% \) and \(\nu =3.87\)MHz. The maximal CDNC to CDC discrepancy is 13.62\% for different values of \(\nu_b\). The CW pump power is around 500\(\mu\)W. Theoretical curves that correspond to the CDNC schemes with different coincidence window \(T_c\) is also plotted in different colors. (b) The ratio between the estimation variance \(\Delta^2 \hat{\eta}_p\) of the CDNC and CDC scheme as a function of noise flux in (a). (c) Experimentally measured (error bar) and theoretically predicted (solid line) estimation variance of CDNC and CDC scheme as a function of source probe flux \(\nu\) with \(\eta_p = 8.53\times 10^{-5} \) and \(\eta_r = 17.4\% \) and \(\nu_b =0.577\)MHz.  The maximal CDNC to CDC discrepancy is 9.21\% for different values of \(\nu\). The CW laser power is swept between zero and 1000\(\mu\)W. Theoretical curves that correspond to the CDNC schemes with different coincidence window \(T_c\) is also plotted in different colors. (d) The ratio between the estimation variance \(\Delta^2 \hat{\eta}_p\) of the CDNC and CDC schemes in (c). The error bar of the plot (a) and (c) is obtained by calculating the standard deviation of the averages of the 3 different groups of 33 independent estimations.}\label{result}
\end{figure*}
The photon pair generation rate is around 4MHz for around 500 \(\mu\)W pump power (\(\simeq10\%\) pump power is coupled into the waveguide).
The bandwidth of the SPDC photon pairs is estimated to be at least 100 nm which corresponds to the intrinsic temporal correlation time \(\Delta t_0\simeq 35\)fs.
The V-polarized reference photons and H-polarized probe photons are separated upon a polarization beam-splitter, after the 783 nm pump light being filtered by a 1200 nm long-pass filter. 
The reference beam is sent directly to a superconducting single-photon detector (the reference detector).
The probe photons are sent toward a weakly reflecting object (a sheet of white paper) \(\approx\)19cm away from the collection optics.
The probe photon that is scattered from the object is collected by the collection optics and detected by a second single-photon detector (the probe detector).
Strong background noise is simulated by shining light from a broadband LED source towards the collection optics.
The noise power is controlled by a tunable attenuator.
The CDC scheme takes place at the same time as the CDNC scheme experiment: instead of blocking reference photons, we extract the CDC scheme data from the CDNC scheme experiment data by discarding all the photon-detection data from the reference detector.
By doing so, the drift of experimental condition between separate CDNC and CDC experiments is eliminated.\\

In what follows we investigate the performance advantage of the CDNC scheme over the CDC scheme under different noise and source flux conditions.
The results are shown in Fig. (\ref{n}) to (\ref{pr}).
The CDNC to CDC enhancement is most pronounced in the high noise \(\nu_b\) and low probe photon flux \(\nu\) regime. The variance ratio between the CDC and CDNC schemes as a function of noise and source power reaches up to 21.36 dB, 26.3 dB respectively, corresponding to a measurement time reduction by a factor of 137 and 430. At the highest environment noise level (right side of the Fig. (\ref{n}) ), the CDNC scheme has performance equal to the CDC scheme with 21.6 dB less noise.   
At the experiment point of highest variance ratio in Fig. (\ref{nr}), the background noise flux detected is 36 dB stronger than the detection rate of the actual probe photons (128 Hz).
Each data point in the figure corresponds to \(99\) independent estimation experiments of \(\tau=300\text{ms}\).
The estimation variance \(\Delta^2\hat{\eta}_p\) is given by the variance of \(33\) independent estimations (the additional factor of  3 contributes to the  error bar in Fig. (\ref{result}) ).
The result of every single estimation is obtained through the maximal likelihood estimation using the experimentally recorded photon detection statistics.
We confirmed that the estimation result of the CDNC and the CDC scheme agrees reasonably when averaged over a large number of independent estimations. The maximal CDNC to CDC discrepancy (the difference between the estimated transmission of CDNC and CDC scheme averaged over \(99\) estimations) is less than 15\%.
The theoretical lower bound curve of the estimation variance in Fig. (\ref{p}) and (\ref{n}) is calculated according to the definition of Fisher information \eqref{FI} and the Cram\'er-Rao's Bound, using the values of parameters \(\mu,\eta_p,\eta_r\) extracted from the photon detection data over a much longer period(30s).
The measured points and the theoretical lower bound are in close agreement, suggesting that the maximal likelihood estimation of \(\eta_p\) approaches the bound of the minimal estimation uncertainty.
It is worth noting that the performance of the CDNC scheme is severely limited by the inefficient transmission \(\eta_r<20\%\) of the reference photons, which is due to the modal mismatch between the mode of the waveguide and the fundamental mode of the fiber.
The performance of the experimental CDNC scheme is also limited by the temporal resolution \(\Delta t \simeq 100\) ps of the detectors. In Fig. (\ref{n})-(\ref{pr}) it is also shown that the performance of the CDNC scheme depends strongly on the temporal coincidence window \(T_c\), which is limited by the effective temporal correlation \(\Delta t_{eff}\).
Since no coincidence detection is involved in the CDC scheme, the temporal resolution of the detector is not relevant to its performance. \\

\begin{figure}[h!]
    \includegraphics[width=0.99\columnwidth]{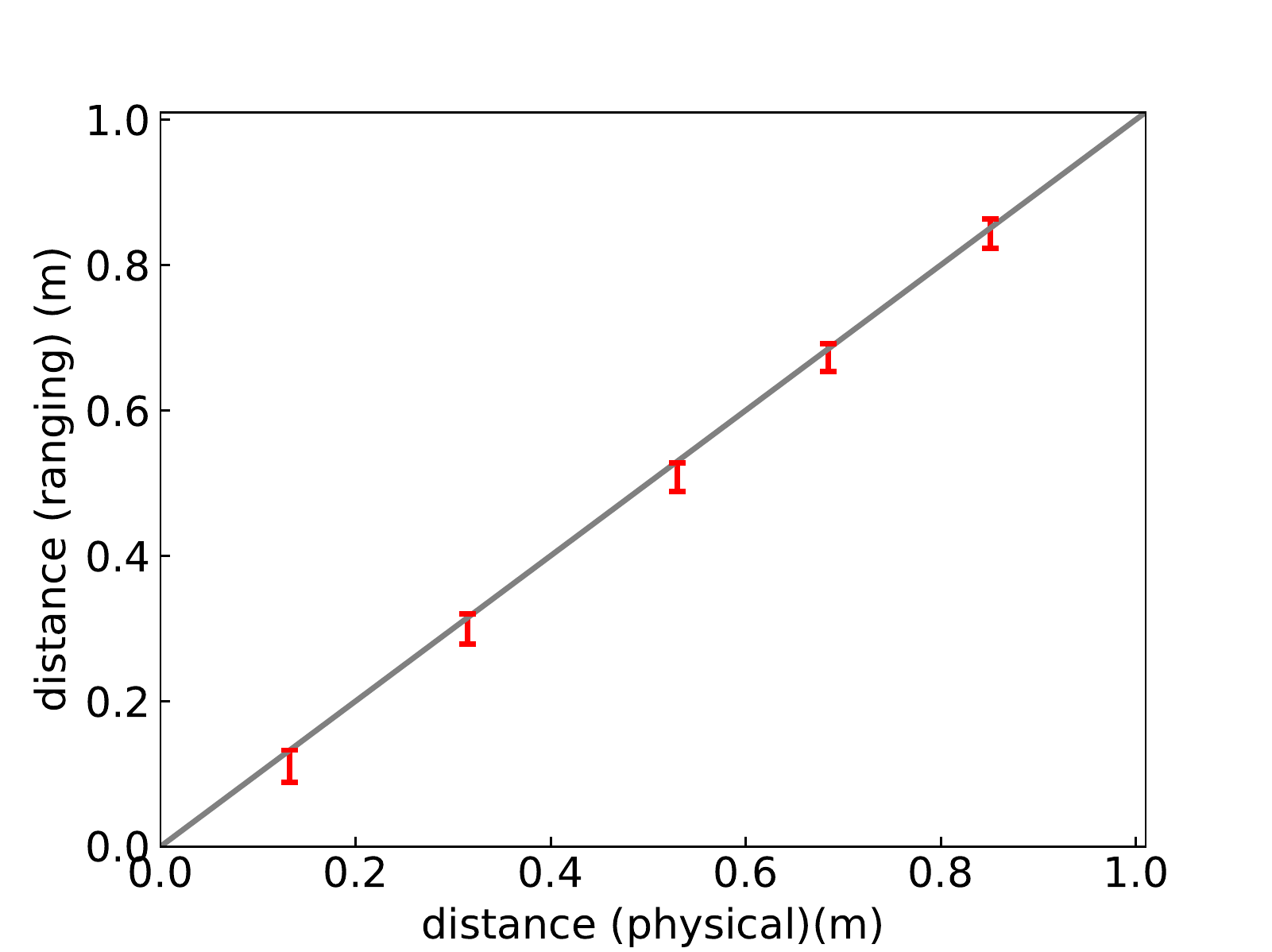}\label{ranging}
    \includegraphics[width=0.99\columnwidth]{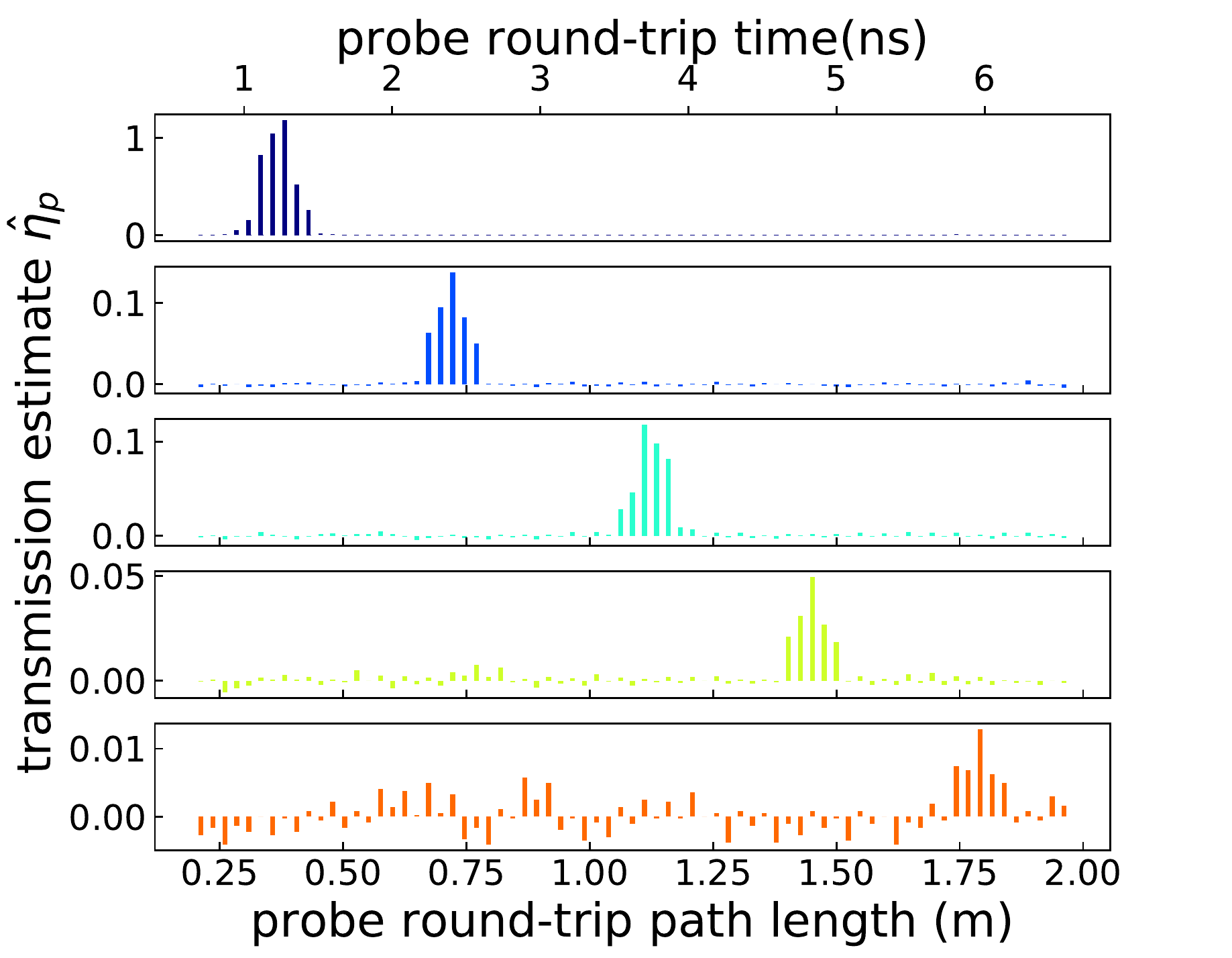}\label{ranging2}
    \caption{ Top: estimated object distance (y-axis) versus the physically measured target distance (x-axis). Red error bar plot: the estimated object distance and its uncertainty. The uncertainty of the distance estimation is taken to be the FWHM of the \(\hat{\eta}_p\) versus \(l_p\) function. Grey solid curve:  the reference curve \(y=x\). In order to increase the ranging range, aluminum foil is used as the object for higher reflectivity. When the target is placed around 0.85 m away, the probe photon detection rate \(\nu\eta_p\) is 26 Hz and the noise photon detection rate \(\nu_b\) is 41.6 KHz. Bottom: estimated \(\hat{\eta}_p\) as a function of probe round-trip path length. The 5-part panel from top to bottom correspond to physical target distance of 13, 32, 53, 68, 85 cm, respectively. }\label{RANGING} 
\end{figure}

We also conduct a ranging experiment based on the CDNC scheme, with the result shown in Fig. (\ref{RANGING}).
The target object is placed at different positions along the optical axis of the collection optics to simulate different target distance.
In order to increase the maximal detection range, we use a piece of aluminum foil that has a higher reflectivity than paper as the object for the ranging experiment.
For each position of the target object, the alignment of the collection optics and output collimator of the probe photons are fine-tuned to maximize the collection efficiency of the probe photons.
To determine the distance of the target object, the transmission estimation \(\hat{\eta}_p\) is calculated for different values of \(l_p\).
The estimated probe path length is taken to be the value of \(l_p\) that maximize the transmission estimation \(\hat{\eta}_p(l_p)\).
The distance of the object is calculated from the estimated probe optical path length based on the geometrical layout of the optical system.
The uncertainty of the target distance is calculated from the full-width half maximum of the function \(\hat{\eta}_p(l_p)\) versus \(l_p\). The \(\hat{\eta}_p\) calculated from the data as a function of \(l_p\) is also shown in Fig. (\ref{RANGING}).
The result shows that even in the presence of dominating background noise, the distance of the target object could be determined with reasonable accuracy of \(\approx\)5 cm, which is limited by the temporal resolution \(\Delta t\) of the detectors. The ranging performance is also closely related to the target detection performance that has been discussed above: the maximal ranging distance is limited to the position where the estimated transmission efficiency \(\hat{\eta}_p\) is comparable to the estimation uncertainty.     

\section{Discussion}
\subsection{Comparison to intensity correlation-based schemes}
The key difference between the presented CDNC scheme and the previously reported intensity correlation-based schemes\cite{Lopaeva:2013}\cite{england2019quantum} is utilizing temporal correlation to substantially and scalably improve the performance of phase-insensitive target detection. In intensity correlation-based schemes, the reference and probe photons are generated in discrete pairs of pulses and only the correlation between the number of the probe and reference photons within each pulse pair is used to enhance the target detection performance. However, there is a fundamental limit in intensity correlation enhancement as the enhancement diminishes as one increases the probe power\cite{england2019quantum}. As a result, there is a trade-off between correlation enhancement and target detection performance. Unlike the intensity correlation-based schemes, the performance enhancement of the CDNC scheme originates from the strong temporal correlation within the non-classical photon pairs and is therefore much more scalable. For the CDNC scheme, the amount of temporal correlation that could be utilized for target detection is characterized by the effective temporal correlation time \(\Delta t_{eff}\). In the current implementation, \(\Delta t_{eff}\) is severely limited by the detector temporal resolution \(\Delta t\simeq100\)ps, which is orders of magnitude larger than the intrinsic correlation time \(\Delta t_0\simeq 35\)fs. However, such detector resolution limit is \textbf{not} fundamental. First of all, it is possible to achieve a better detector temporal resolution with improved single-photon detection technology. For example, if the detector temporal resolution could be improved to \(\Delta t = 15\)ps\cite{esmaeil2017single}, the noise resilience of the CDNC scheme (which is inversely proportional to the effective temporal correlation time \(\Delta t_{eff}\)) could be further enhanced by a factor of 7. More importantly, even with commercially available detectors (\(\Delta t \simeq 50\)ps), it is still possible to fully exploit the strong intrinsic temporal correlation via novel photon detection techniques. In \supp we give an example of such technique that is based on dispersion cancellation. It is shown that the effective detector temporal resolution could be improved from 50ps to \(\simeq\)500fs by adding equal amount but opposite sign of group velocity dispersion (5km of the standard 1550nm single-mode fiber) to both the probe and the reference photons. The intrinsic temporal correlation time \(\Delta t_0\), which is the fundamental limiting factor of the CDNC scheme performance, could be improved with different SPDC waveguide structure designs \cite{Abolghasem:2009}. 
\subsection{Covert operation with the CDNC scheme}
The previous sections have shown the significant performance advantages of the CDNC scheme over the CDC scheme in a lossy and noisy environment as quantified by the transmission estimation/Fisher information criterion.
However, it is important to note that the performance enhancement of the CDNC scheme is based on a key assumption we made about the property of the background noise: namely we assumed that the background noise photons are indistinguishable from the probe photons in terms of their spectral and temporal distribution, i.e., CW and broadband. This assumption is valid for the CDNC scheme for the following reasons. First, it is reasonable to assume that the environmental noise is CW, the same as the SPDC probe light. Second, the spectrum of the SPDC probe photons could be tailored to match that of the noise photons through different waveguide structure designs\cite{Abolghasem:2009}. Third, in some covert detection application, it may be desirable to deliberately inject noise power that is indistinguishable from the probe light such that the signal to noise ratio for the un-authorized receivers is minimized.
However, for the CDC schemes, it is possible to reduce the in-band noise power through temporal or spectral shaping of the probe light.
For example, one could utilize pulsed or monochromatic light instead of CW broadband light to probe the target object, such that the spectral-temporal overlap between the probe and the noise light is minimized. (see \supp for the detailed analysis)
Nevertheless, the temporal or spectral concentration of optical energy will increase the distinguishability between the probe photon and the noise photon, hence increase the visibility of the target detection channel and its vulnerability to active attacks.
Although classical scrambling techniques such as frequency scrambling could help hide the probe light under the disguise of background noise, such indistinguishability between probe and noise photons is not guaranteed by fundamental principles of physics.
The CDNC scheme, on the other hand, could offer unconditional indistinguishability between the probe and the background noise photon. This is because each probe photon in the CDNC scheme is generated at a truly random time and frequency\cite{xu2016experimental}.  
From this standpoint, the CDNC scheme is an analogue of the quantum noise radar\cite{chang2019quantum} where the probe and the reference signal are completely random but correlated.
Covert ranging is one example of the stealth operation property of the CDNC scheme.
Classical optical ranging systems typically use time-variant optical signal such as pulsed light to probe the target object.
This is because the time-invariant back-reflected signal contains little information about the target distance.
However, time-variant probe light that is distinguishable from the CW background noise will make the ranging channel visible, hence vulnerable to active attacks.
On the other hand, in the CDNC scheme, the probe light generated is time-invariant and indistinguishable from the background noise for an unauthorized receiver.
Yet the distance of the object to be detected could still be calculated from detected temporal correlation statistics.

\section{Conclusion}
We demonstrated a phase-insensitive target detection scheme enhanced by non-classical light generated in a semiconductor chip source.
We quantified the performance of both the enhanced and classical target detection schemes with the estimation uncertainty and the Fisher information of the probe photon transmission.
We showed that the strong temporal correlation of non-classical photon pairs could be utilized to enhance the target detection performance by up to 26.3 dB in a lossy and noisy environment, which is equivalent to a reduction of target detection time by a factor of 430. 
Such performance enhancement is highly scalable with improved photon detection temporal resolution or improved photon detection techniques.
We also showed that this enhanced scheme could be used for ranging with CW source in a lossy and noisy environment. 
Due to the enhanced performance and the phase-insensitive nature, this non-classical source enhanced target detection scheme with semiconductor chip sources could find its application in many real-world sensing scenarios. \\

See \supp for supporting content.
\bibliography{References.bib}

\end{document}